\documentclass[twocolumn,showpacs,preprintnumbers,amsmath,amssymb,eqsecnum]{revtex4}
\usepackage{dcolumn}
\usepackage{bm}
\usepackage[dvips]{graphicx}
\usepackage{wrapfig}
\usepackage{psfrag}
\usepackage{color}
\begin{document}

\def\sh{\mathop{\rm sh}\nolimits}
\def\ch{\mathop{\rm ch}\nolimits}
\def\var{\mathop{\rm var}}\def\exp{\mathop{\rm exp}\nolimits}
\def\Re{\mathop{\rm Re}\nolimits}
\def\Sp{\mathop{\rm Sp}\nolimits}
\def\kp{\mathop{\text{\ae}}\nolimits}
\def\bk{{\bf {k}}}
\def\bp{{\bf {p}}}
\def\bq{{\bf {q}}}
\def\lra{\mathop{\longrightarrow}}
\def\Const{\mathop{\rm Const}\nolimits}
\def\sh{\mathop{\rm sh}\nolimits}
\def\ch{\mathop{\rm ch}\nolimits}
\def\var{\mathop{\rm var}}
\def\mK{\mathop{{\mathfrak {K}}}\nolimits}
\def\mR{\mathop{{\mathfrak {R}}}\nolimits}
\def\mv{\mathop{{\mathfrak {v}}}\nolimits}
\def\mV{\mathop{{\mathfrak {V}}}\nolimits}
\def\mD{\mathop{{\mathfrak {D}}}\nolimits}
\def\mN{\mathop{{\mathfrak {N}}}\nolimits}
\def\mS{\mathop{{\mathfrak {S}}}\nolimits}

\newcommand\ve[1]{{\mathbf{#1}}}

\def\Re{\mbox {Re}}
\newcommand{\Z}{\mathbb{Z}}
\newcommand{\R}{\mathbb{R}}
\def\mK{\mathop{{\mathfrak {K}}}\nolimits}
\def\mk{\mathop{{\mathfrak {k}}}\nolimits}
\def\mR{\mathop{{\mathfrak {R}}}\nolimits}
\def\mv{\mathop{{\mathfrak {v}}}\nolimits}
\def\mV{\mathop{{\mathfrak {V}}}\nolimits}
\def\mD{\mathop{{\mathfrak {D}}}\nolimits}
\def\mN{\mathop{{\mathfrak {N}}}\nolimits}
\def\ml{\mathop{{\mathfrak {l}}}\nolimits}
\def\mf{\mathop{{\mathfrak {f}}}\nolimits}
\newcommand{\ccm}{{\cal M}}
\newcommand{\cE}{{\cal E}}
\newcommand{\cV}{{\cal V}}
\newcommand{\cI}{{\cal I}}
\newcommand{\cR}{{\cal R}}
\newcommand{\cK}{{\cal K}}
\newcommand{\cH}{{\cal H}}
\newcommand{\cW}{{\cal W}}

\def\br{\mathop{{\bf {r}}}\nolimits}
\def\bS{\mathop{{\bf {S}}}\nolimits}
\def\bA{\mathop{{\bf {A}}}\nolimits}
\def\bJ{\mathop{{\bf {J}}}\nolimits}
\def\bn{\mathop{{\bf {n}}}\nolimits}
\def\bg{\mathop{{\bf {g}}}\nolimits}
\def\bv{\mathop{{\bf {v}}}\nolimits}
\def\be{\mathop{{\bf {e}}}\nolimits}
\def\bp{\mathop{{\bf {p}}}\nolimits}
\def\bz{\mathop{{\bf {z}}}\nolimits}
\def\bbf{\mathop{{\bf {f}}}\nolimits}
\def\bb{\mathop{{\bf {b}}}\nolimits}
\def\ba{\mathop{{\bf {a}}}\nolimits}
\def\bx{\mathop{{\bf {x}}}\nolimits}
\def\by{\mathop{{\bf {y}}}\nolimits}
\def\br{\mathop{{\bf {r}}}\nolimits}
\def\bs{\mathop{{\bf {s}}}\nolimits}
\def\bH{\mathop{{\bf {H}}}\nolimits}
\def\bk{\mathop{{\bf {k}}}\nolimits}
\def\be{\mathop{{\bf {e}}}\nolimits}
\def\bnul{\mathop{{\bf {0}}}\nolimits}
\def\bq{{\bf {q}}}

\newcommand{\oV}{\overline{V}}
\newcommand{\vkp}{\varkappa}
\newcommand{\os}{\overline{s}}
\newcommand{\opsi}{\overline{\psi}}
\newcommand{\ov}{\overline{v}}
\newcommand{\oW}{\overline{W}}
\newcommand{\oPhi}{\overline{\Phi}}
\newcommand{\C}{\mathbb{C}}

\def\mI{\mathop{{\mathfrak {I}}}\nolimits}
\def\mA{\mathop{{\mathfrak {A}}}\nolimits}

\def\st{\mathop{\rm st}\nolimits}
\def\tr{\mathop{\rm tr}\nolimits}
\def\sign{\mathop{\rm sign}\nolimits}
\def\const{\mathop{\rm const}\nolimits}
\def\O{\mathop{\rm O}\nolimits}
\def\Spin{\mathop{\rm Spin}\nolimits}
\def\exp{\mathop{\rm exp}\nolimits}
\def\SU{\mathop{\rm SU}\nolimits}
\def\mU{\mathop{{\mathfrak {U}}}\nolimits}
\newcommand{\cU}{{\cal U}}
\newcommand{\cD}{{\cal D}}

\def\mI{\mathop{{\mathfrak {I}}}\nolimits}
\def\mA{\mathop{{\mathfrak {A}}}\nolimits}
\def\mU{\mathop{{\mathfrak {U}}}\nolimits}

\def\st{\mathop{\rm st}\nolimits}
\def\tr{\mathop{\rm tr}\nolimits}
\def\sign{\mathop{\rm sign}\nolimits}
\def\d{\mathop{\mathrm d}\nolimits}
\def\d{\mathrm{d}}
\def\const{\mathop{\rm const}\nolimits}
\def\O{\mathop{\rm O}\nolimits}
\def\Spin{\mathop{\rm Spin}\nolimits}
\def\exp{\mathop{\rm exp}\nolimits}

\title{Fermion zero mode Associated with instantonlike self-dual solution to lattice Euclidean gravity}

\author {S.N. Vergeles\vspace*{4mm}\footnote{{e-mail:vergeles@itp.ac.ru}}}

\affiliation{Landau Institute for Theoretical Physics,
Russian Academy of Sciences,
Chernogolovka, Moscow region, 142432 Russia \linebreak
and   \linebreak
Moscow Institute of Physics and Technology, Department
of Theoretical Physics, Dolgoprudnyj, Moskow region,
141707 Russia}

\begin{abstract} We prove the existence of lattice fermion zero mode associated with
self-dual lattice gravity solution.
\end{abstract}

\pacs{11.15.-q, 11.15.Ha}

\maketitle

\section{Introduction}

It is  known that in continuum relativistic gauge theories coupled with fermions  some of the currents conserved in classical mechanics become non-conserved ones in quantum mechanics due to vacuum quantum fluctuations: the divergences of some currents are equal to a certain local functions of
the gauge field called "anomaly" which are generally not zero.

Consider, for example, 4D Euclidean Yang-Mills theory. The following formulae are well known:
\begin{gather}
\partial_{\mu}\left(i\Psi^{\dag}\gamma^{\mu}\gamma^5\Psi\right)=
\nonumber \\
=-2\sum_{N:\,|\epsilon_N|<\Lambda\longrightarrow\infty}\Psi^{\dag}_N\gamma^5\Psi_N=
\frac{e^2}{16\pi^2}\varepsilon^{\mu\nu\lambda\rho}\tr F_{\mu\nu}F_{\lambda\rho},
\label{intr10}
\end{gather}
\begin{gather}
i\gamma^{\mu}\nabla_{\mu}\Psi_N=\epsilon_N\Psi_N,
\nonumber \\
\nabla_{\mu}=\partial_{\mu}+ieA_{\mu},  \quad [\nabla_{\mu},\,\nabla_{\nu}]=-ieF_{\mu\nu}.
\nonumber
\end{gather}
Let's integrate the last equation in (\ref{intr10}) over space. We obtain
\begin{gather}
\sum_{N_0:\,\epsilon_{N_0}=0}\int\d^{(4)}x\,\Psi^{\dag}_{N_0}\gamma^5\Psi_{N_0}=
\nonumber \\
=-\frac{e^2}{32\pi^2}\varepsilon^{\mu\nu\lambda\rho}
\int\d^{(4)}x\,\tr F_{\mu\nu}F_{\lambda\rho}=q.
\label{intr20}
\end{gather}
since the modes $\Psi_N$ and $\gamma^5\Psi_N$ are mutually orthogonal for $\epsilon_N\neq0$. Here $q=0,\,\pm1,\ldots$ is a topological charge of the Yang-Mills field instanton. Now the Atiyah-Singer index theorem is obtained if we substitute for $\gamma^5$ its decomposition $\gamma^5\equiv(1/2)\left(1+\gamma^5\right)-(1/2)\left(1-\gamma^5\right)$ into left hand side of Eq. (\ref{intr20}):
\begin{gather}
n_+-n_-=q.
\label{intr30}
\end{gather}
Here $n_+$ ($n_-$) is the number of right (left) fermion zero modes associated with instanton with
the topological charge $q$.

It is seen from this consideration that the existence of fermion zero modes associated with instanton in Yang-Mills theory is provided by the existence of anomaly in divergence of the corresponding fermion axial current (\ref{intr10}).

But the problem of the anomalies and their connection with fermion zero modes in lattice gauge
theories is qualitatively more complicated one (see \cite{1}). Note that all lattice theories under consideration
possess the common fundamental property: lattice theories transform into corresponding continuum relativistic theories at the naive long-wavelength limit.

Let's consider firstly the Yang-Mills instanton in a lattice theory. The configuration of the Yang-Mills instanton field is smooth at each regions of space-time. This property is very important for validity of the second  equality in (\ref{intr10}). Indeed, this equality is obtained correctly only for long-wavelength (as compared to fermion fields wavelengths) gauge fields. Therefore, the second  equality in (\ref{intr10}) is valid in the lattice theories in the naive long-wavelength limit. This property of the Yang-Mills theory implies very important physical  consequences. In particular, it follows from here that the irregular ultrashort (doubled) fermion quanta with low energy also exist \cite{1} besides soft regular long-wavelength fermion quanta.

The gravity equivalent of the second equality in (\ref{intr10}) fails since the gravitational instanton field configuration is singular near the centre of the instanton (see Section III) in lattice gravity theory. Therefore a proof of the 
existence of the lattice fermion zero mode associated with the instanton would be significantly different from
the above exposed method. We use here the method that has been successful in solving lattice pure gravity self-dual equations with given boundary conditions \cite{2} (lattice instanton).

The method makes it possible to establish the main result of the paper: a proof of the existence of fermion zero mode associated with a lattice gravitational instanton. Here we emphasize that the lattice approach developed in \cite{2}
and used here can not be extended to usual continuum field theories since any finite space-time region contains innumerable set of variables in such theories.

The following extremely important point is in order.
The Eguchi-Hanson continuous solution (see \cite{3}, \cite{4}) is valid for the
manifold $M$ with boundary as $r\longrightarrow\infty$ which is the cotangent bundle of the complex plane, $P_1(\C)\approx S^2$:
\begin{gather}
M=T^*\left(P_1(\C)\right),
\nonumber \\
\partial M=SO(3)=S^3/Z_2.
\label{intr30}
\end{gather}
The manifold $M$ is smooth.
On the other hand, the discrete analogue of the Eguchi-Hanson solution \cite{2} and Dirac zero mode exist on a triangulation of manifold $\R^4$ which can be considered as $S^3$ of extra-large $r$ including its interior. This  triangulation
is designated as $\mK'$ (see Section 3), $\partial\mK'\approx S^3$. Evidently, the topologies of the
manifold $M$ and simplicial complex $\mK'$ are different.

The organization of the paper is as follows.
In Sections II  and III the early obtained results which are necessary here are shortly outlined:
the definition of lattice gravity theory and self-dual solution on the lattice.
In Section IV the asymptotic behaviour on a long-wavelength limit of fermion zero mode is studied.
In Section V the existence of lattice fermion zero mode associated with self-dual solution is proved
with the help of the method used for proof of existence lattice self-dual solution \cite{2}.

\section{The lattice gravity model}

Let's introduce some designations:
\begin{gather}
\gamma^{\alpha}=
 \left(
\begin{array}{cc}
0 & i\sigma^{\alpha}  \\
-i\sigma^{\alpha} & 0 \\
\end{array} \right),
\quad \gamma^4=
 \left(
\begin{array}{cc}
0 & 1  \\
1 & 0 \\
\end{array} \right),
\quad \alpha=1,2,3,
\nonumber \\
\gamma^5\equiv\gamma^1\gamma^2\gamma^3\gamma^4=
 \left(
\begin{array}{cc}
-1 & 0  \\
0 & 1 \\
\end{array} \right),
\nonumber \\
\sigma^{ab}=\frac{1}{4}[\gamma^a,\gamma^b], \quad a,\,b,\ldots=1,2,3,4,
\nonumber \\
\sigma^{\alpha 4}= \frac{i}{2}\left(
\begin{array}{cc}
\sigma^{\alpha} & 0  \\
0 & -\sigma^{\alpha}  \\
\end{array} \right),
\quad
\sigma^{\alpha\beta}= \frac{i\varepsilon_{\alpha\beta\gamma}}{2}\left(
\begin{array}{cc}
\sigma^{\gamma} & 0  \\
0 & \sigma^{\gamma}  \\
\end{array} \right),
\label{discr10}
\end{gather}
$\sigma^{\alpha}$ are Pauli matrices.

It is necessary to sketch out the model of lattice gravity which is used here. A detailed description of the model is given in \cite{1}, \cite{5}, \cite{6}.

The orientable 4-dimensional simplicial complex and its vertices are designated as $\mK$ and
$a_{\cV}$, the indices ${\cV}=1,2,\dots,\,{\mN}\rightarrow\infty$ and ${\cW}$ enumerate the vertices
and 4-simplices, correspondingly. We assume here that $\mK\approx{\boldsymbol{\R^4}}$ in a topological sense. It is necessary to use
the local enumeration of the vertices $a_{\cV}$ attached to a given
4-simplex: the all five vertices of a 4-simplex with index ${\cW}$
are enumerated as $a_{{\cW}i}$, $i=1,2,3,4,5$. The later notations with extra index  ${\cW}$
indicate that the corresponding quantities belong to the
4-simplex with index ${\cW}$. The Levi-Civita symbol with in pairs
different indexes $\varepsilon_{{\cW}ijklm}=\pm 1$ depending on
whether the order of vertices
$s^4_{\cW}=a_{{\cW}i}a_{{\cW}j}a_{{\cW}k}a_{{\cW}l}a_{{\cW}m}$ defines the
positive or negative orientation of 4-simplex $s^4_{\cW}$.

An element of the group $\Spin(4)$ and an element of the Clifford algebra
\begin{gather}
\Omega_{{\cW}ij}=\Omega^{-1}_{{\cW}ji}=\exp\left(\omega_{{\cW}ij}\right), \quad
\omega_{{\cW}ij}\equiv\frac{1}{2}\sigma^{ab}\omega^{ab}_{{\cW}ij},
\nonumber \\
\hat{e}_{{\cW}ij}=\hat{e}_{{\cW}ij}^{\dag}\equiv e^a_{{\cW}ij}\gamma^a\equiv-\Omega_{{\cW}ij}\hat{e}_{{\cW}ji}\Omega_{{\cW}ij}^{-1}.
\label{discr20}
\end{gather}
are assigned for each oriented 1-simplex $a_{{\cW}i}a_{{\cW}j}$.
The Dirac spinors $\Psi_{\cV}$ and $\Psi^{\dag}_{\cV}$,
each of whose components assumes
values in a complex Grassman algebra, are assigned to each vertex $a_{\cV}$.
In the case of Euclidean
signature, the spinors $\Psi_{\cV}$ and $\Psi^{\dag}_{\cV}$ are
independent variables and are interchanged under the Hermitian
conjugation.

Thus, the used representation  realizes automatically the separation of a total gauge group into two sub-group:
$\Spin(4)\approx\Spin(4)_{(+)}\otimes\Spin(4)_{(-)}$. For example
\begin{gather}
\frac12\sigma^{ab}\omega^{ab}_{{\cW}ij}=
 \frac{i\sigma^{\alpha}}{2} \left(
\begin{array}{cc}
\omega^{\alpha}_{(+){\cW}ij} & 0  \\
0 & \omega^{\alpha}_{(-){\cW}ij} \\
\end{array} \right),
\nonumber \\
\omega^{\alpha}_{(\pm){\cW}ij}\equiv \left\{\pm\omega^{\alpha4}_{{\cW}ij}+\frac12\varepsilon_{\alpha\beta\gamma}
\omega^{\beta\gamma}_{{\cW}ij}\right\}.
\label{discr30}
\end{gather}
The underwritten lattice instanton solution and fermion
zero mode are described in terms of the sub-group $\Spin(4)_{(+)}$.

The considered lattice action has the form
\begin{gather}
\mA=\mA_g+\mA_{\Psi},
\label{discr40}
\end{gather}
\begin{gather}
\mA_g=-\frac{1}{ 5\cdot
24\cdot2\cdot l^2_P}\sum_{\cW}\sum_{i,j,k,l,m}\varepsilon_{{\cW}ijklm}\times
\nonumber \\
\times\tr\,\gamma^5
\Omega_{{\cW}mi}\Omega_{{\cW}ij}\Omega_{{\cW}jm}
\hat{e}_{{\cW}mk}\hat{e}_{{\cW}ml},
\label{discr50}
\end{gather}
\begin{gather}
\mA_{\Psi}=-\frac{1}{5\cdot24^2}\sum_{\cW}\sum_{i,j,k,l,m}\varepsilon_{{\cW}ijklm}\times
\nonumber \\
\times\tr\,\gamma^5 \hat{\Theta}_{{\cW}mi}
\hat{e}_{{\cW}mj}\hat{e}_{{\cW}mk}\hat{e}_{{\cW}ml},
\nonumber \\
\hat{\Theta}_{{\cW}ij}=
\frac{i}{2}\gamma^a\left(\Psi^{\dag}_{{\cW}i}\gamma^a
\Omega_{{\cW}ij}\Psi_{{\cW}j}-\Psi^{\dag}_{{\cW}j}\Omega_{{\cW}ji}\gamma^a\Psi_{{\cW}i}\right)\equiv
\nonumber \\
\equiv\Theta^a_{{\cW}ij}\gamma^a.
\label{discr60}
\end{gather}
This action is invariant relative to the gauge transformations
\begin{gather}
\tilde{\Omega}_{{\cW}ij}=S_{{\cW}i}\Omega_{{\cW}ij}S^{-1}_{{\cW}j},
\quad
\tilde{e}_{{\cW}ij}=S_{{\cW}i}\,e_{{\cW}ij}\,S^{-1}_{{\cW}i},
\nonumber \\
\tilde{\Psi}_{{\cW}i}=S_{{\cW}i}\,\Psi_{A\,i},  \quad
\tilde{\Psi^{\dag}}_{{\cW}i}=\Psi^{\dag}_{{\cW}i}\,S^{-1}_{{\cW}i}
\nonumber \\
 S_{{\cW}i}\in\Spin(4).
\label{discr70}
\end{gather}

The action (\ref{discr40}) reduces to the continuum action of gravity in a four-dimensional Euclidean space in the limit of slowly varying fields, minimally connected with a Dirac field.

Consider a certain  $4D$ sub-complex of complex $\mK$ with the trivial topology of four-dimensional disk.
Realize geometrically this sub-complex in $\R^4$.  Suppose that the
geometric realization is an almost smooth four-dimensional
surface {\footnote{Here, by an almost smooth surface, we
mean a piecewise smooth surface consisting of flat
four-dimensional simplices, such that the angles between
adjacent 4-simplices tend to zero and the sizes of these
simplices are commensurable.}}. Thus each vertex of the sub-complex  acquires
the coordinates $x^{\mu}$  which are the coordinates of the vertex image in $\R^4$:
\begin{gather}
x^{\mu}_{{\cW}i}=x^{\mu}_{\cV}\equiv x^{\mu}(a_{{\cW}i})\equiv x^{\mu}(a_{\cV}),
 \qquad \ \mu=1,\,2,\,3,\,4
\label{discr80}
\end{gather}
We stress that these coordinates
are defined only by their vertices rather than by the higher dimension simplices
to which these vertices belong; moreover, the correspondence between the vertices
from the considered subset and the coordinates (\ref{discr80}) is one-to-one.

The four vectors
\begin{gather}
\d x^{\mu}_{{\cW}ji}\equiv x^{\mu}_{{\cW}i}-x^{\mu}_{{\cW}j},
 \quad i=1,\,2,\,3,\,4
\label{discr90}
\end{gather}
are linearly independent and
\begin{gather}
\left\vert
\begin{array}{llll}
\d x^1_{{\cW}m1} \ & \ \d x^2_{{\cW}m1} \ & \ldots & \ \d x^4_{{\cW}m1}\\
\ldots  & \ldots  &  \ldots  & \ldots \\
\d x^1_{{\cW}m4} \ & \ \d x^2_{{\cW}m4} \ & \ldots & \ \d x^4_{{\cW}m4}
\end{array}\right\vert\gtrless0,
\label{discr100}
\end{gather}
depending on whether the frame $\big(X^{\cW}_{m\,1},\,\ldots\,,\,X^{\cW}_{m\,4}\,\big)$ is positively or negatively oriented. Here, the differentials of coordinates (\ref{discr90}) correspond to one-dimensional simplices $a_{{\cW}j}a_{{\cW}i}$, so that, if the vertex $a_{{\cW}j}$ has coordinates $x^{\mu}_{{\cW}j}$, then the vertex $a_{{\cW}i}$ has the coordinates $x^{\mu}_{{\cW}j}+\d x^{\mu}_{{\cW}ji}$.

In the continuous limit, the holonomy group elements (\ref{discr20}) are
close to the identity element, so that the quantities $\omega^{ab}_{ij}$
tend to zero being of the order of $O(\d x^{\mu})$.
Thus one can consider the following system of equation for $\omega_{{\cW}m\mu}$
\begin{gather}
\omega_{{\cW}m\mu}\,\d x^{\mu}_{{\cW}mi}=\omega_{{\cW}mi},  \quad
i=1,\,2,\,3,\,4\,.
\label{discr110}
\end{gather}
In this system of linear equation, the indices ${\cW}$ and $m$ are
fixed, the summation is carried out over the index $\mu$, and
index runs over all its values. Since the determinant
(\ref{discr100}) is nonzero, the quantities $\omega_{{\cW}m\mu}$
are defined uniquely. Suppose that a one-dimensional simplex
$X^{\cW}_{m\,i}$ belongs to four-dimensional simplices with indices
${\cW}_1,\,{\cW}_2,\,\ldots\,,\,{\cW}_r$. Introduce the quantity
\begin{gather}
\omega_{\mu}\left(\frac{1}{2}\,(x_{{\cW}m}+
x_{{\cW}i})\,\right)\equiv\frac{1}{r}\,
\bigg\{\omega_{{\cW}_1m\mu}+\,\ldots\,+\omega_{{\cW}_rm\mu}\,\bigg\}\,,
\label{discr120}
\end{gather}
which is assumed to be related to the midpoint of the segment
$[x^{\mu}_{{\cW}m},\,x^{\mu}_{{\cW}i}\,]$. Recall that the coordinates
$x^{\mu}_{{\cW}i}$ as well as the differentials (\ref{discr90})
depend only on vertices but not on the higher dimensional
simplices to which these vertices belong. According to the
definition, we have the following chain of equalities
\begin{gather}
\omega_{{\cW}_1\,mi}=\omega_{{\cW}_2\,mi}= \,\ldots\,=\omega_{{\cW}_r\,mi}\,.
\label{discr130}
\end{gather}
It follows from (\ref{discr90}) and
(\ref{discr110})--(\ref{discr130}) that
\begin{gather}
\omega_{\mu}\left(x_{{\cW}m}+ \frac{1}{2}\,\d x_{{\cW}mi}\,\right)\,\d
x^{\mu}_{{\cW}mi}=\omega_{{\cW}mi}  \,.
\label{discr140}
\end{gather}
The value of the field element $\omega_{\mu}$ in (\ref{discr140}) is uniquely defined by the corresponding
one-dimensional simplex.

Next, we assume that the fields $\omega_{\mu}$ smoothly depend on
the points belonging to the geometric realization of each
four-dimensional simplex. In this case, the following formula is
valid up to $O\big((\d x)^2\big)$ inclusive
\begin{gather}
\Omega_{{\cW}mi}\,\Omega_{{\cW}ij}\,\Omega_{{\cW}jm}=
\nonumber \\
=\exp\left[\frac{1}{2}\,\mR_{\mu\nu}(x_{{\cW}m})\d x^{\mu}_{{\cW}mi}\, \d
x^{\nu}_{{\cW}mj}\,\right]\,,
\label{discr150}
\end{gather}
where
\begin{gather}
\mR_{\mu\nu}=\partial_{\mu}\omega_{\nu}-\partial_{\nu}\omega_{\mu}+
[\omega_{\mu},\,\omega_{\nu}\,]\equiv
\frac12\sigma^{ab}\mR^{ab}_{\mu\nu},
\nonumber \\
\mR^{ab}\equiv \mR^{ab}_{\mu\nu}\d x^{\mu}\wedge\d x^{\nu}.
\label{discr160}
\end{gather}

In exact analogy with (\ref{discr110}), let us write out the following relations
for a tetrad field without explanations
\begin{gather}
\hat{e}_{{\cW}m\mu}\,\d x^{\mu}_{{\cW}mi}=\hat{e}_{{\cW}mi}\longrightarrow
e^a=e^a_{\mu}\d x^{\mu}.
\label{discr170}
\end{gather}

Using (\ref{discr20}), (\ref{discr90}) and (\ref{discr110}), we can rewrite the 1-form
(\ref{discr60}) as
\begin{gather}
\hat{\Theta}_{{\cW}ij}=\gamma^a\,\frac{i}{2}\,\left[
\Psi^{\dag}\gamma^a\,{\cal D}_{\mu}\,\Psi-
\left({\cal D}_{\mu}\,\Psi\right)^{\dag}\gamma^a\,\Psi
\right]\d x^{\mu}_{A\,ij}\equiv
\nonumber \\
\equiv\Theta^a\gamma^a,
\label{discr180}
\end{gather}
to within $O(\d x)$; here,
\begin{gather}
{\cal D}_{\mu}\,\Psi=\partial_{\mu}
\Psi+\omega_{\mu}\Psi,
\quad
\Psi=
\left(
\begin{array}{c}
\phi \\
\eta \\
\end{array} \right).
\label{discr190}
\end{gather}
and the smooth field $\Psi(x)$ takes the values $\Psi(x_{{\cW}i})=\Psi_{{\cW}i}$.

Applying formulas (\ref{discr150})--(\ref{discr180}) to the discrete action (\ref{discr40}) and
changing the summation to integration we obtain in the continuum limit the well known
gravity action:
\begin{gather}
\mA=\int\,\varepsilon_{abcd}\,\left\{-\frac{1}{l^2_P}\mR^{ab}\wedge e^c\wedge e^d-\right.
\nonumber \\
\left.-\frac{1}{6}\,\Theta^a\wedge e^b\ \wedge e^c\wedge e^d
\right\}.
\label{discr200}
\end{gather}

Thus, in the naive continuum limit, the action
(\ref{discr40}) proves to be equal to the gravity action in the Palatini form  minimally coupled to a Dirac field
with Euclidean signature.

Another way of constructing Dirac fermions on simplicial complexes is  stated in \cite{7}.

\section{The lattice gravitational instanton}

Let's consider firstly the instanton field configuration far apart from the instanton centre
where the continuous limit is valid (Eguchi-Hanson solution).

The following designations
\begin{widetext}
\begin{gather}
\sigma^a\equiv\left(\sigma^1,\,\sigma^2,\,\sigma^3,\,i\right),  \quad
\d x^{\mu}=\left(
\begin{array}{cccc}
\d \theta, & \d \varphi, & \d \psi,  & \d r \\
\end{array} \right), \quad a=1,2,3,4,
\nonumber \\
\partial_{\mu}\equiv
\left(
\begin{array}{c}
\partial_{\theta} \\
\partial_{\varphi} \\
\partial_{\psi} \\
\partial_r \\
\end{array} \right), \quad
\varsigma^1\equiv
\left(
\begin{array}{c}
\sin\psi \\
-\sin\theta\cos\psi \\
0 \\
0 \\
\end{array} \right), \quad
\varsigma^2\equiv
\left(
\begin{array}{c}
\cos\psi \\
\sin\theta\sin\psi \\
0 \\
0 \\
\end{array} \right), \quad
\varsigma^3\equiv
\left(
\begin{array}{c}
0 \\
-\cos\theta \\
-1 \\
0 \\
\end{array} \right), \quad
\varsigma^4\equiv
\left(
\begin{array}{c}
0 \\
0 \\
0 \\
1 \\
\end{array} \right).
\label{inst10}
\end{gather}
\end{widetext}
for the  row and column matrices are used.
We have
\begin{gather}
\d x^{\mu}e_{\mu}^a=\left(
\begin{array}{cccc}
\d \theta, & \d \varphi, & \d \psi,  & \d r \\
\end{array} \right)\times
\nonumber \\
\times\left(
\begin{array}{cccc}
\frac12r\varsigma^1, & \frac12r\varsigma^2, & \frac12rg\varsigma^3,  & g^{-1}\varsigma^4 \\
\end{array} \right),
\nonumber \\
g=\sqrt{1-\frac{a^4}{r^4}}.
\label{inst20}
\end{gather}
for Eguchi-Hanson self-dual solution to continuous Euclidean
Gravity \cite{3}-\cite{4}.

The $4\times4$ matrix  which is inverse to that in (\ref{inst20}) is of the form
%\begin{widetext}
\begin{gather}
e^{\mu}_a=2\times
\nonumber \\
\times\left(\!\!
\begin{array}{cccc}
r^{-1}\sin\psi  &   -(r\sin\theta)^{-1}\cos\psi  & r^{-1}\cot\theta\cos\psi & 0 \\
r^{-1}\cos\psi   &  (r\sin\theta)^{-1}\sin\psi  & -r^{-1}\cot\theta\sin\psi & 0  \\
0 & 0  &   -(rg)^{-1} &  0 \\
0  &  0  & 0  &  g/2  \\
\end{array}\!\! \right)\!.
\label{inst30}
\end{gather}
%\end{widetext}

For the instanton gravitational field we have
\begin{widetext}
\begin{gather}
\frac12\sigma^{ab}\omega_{\mu}^{ab}=
 \frac{i\sigma^{\alpha}}{2} \left(
\begin{array}{cc}
\omega^{\alpha}_{(+)\mu} & 0  \\
0 & 0 \\
\end{array} \right),
\label{inst40}
\end{gather}
\begin{gather}
\frac{i}{2}\d x^{\mu}\left(\omega^{\alpha}_{(+)\mu}\sigma^{\alpha}\right)=
\frac{i}{2}\left(
\begin{array}{cccc}
\d \theta, & \d \varphi, & \d \psi,  & \d r \\
\end{array} \right)\bigg(
g\varsigma^1\sigma^1+g\varsigma^2\sigma^2+(2-g^2)\varsigma^3\sigma^3 \bigg).
\label{inst50}
\end{gather}
\end{widetext}

Now we describe the lattice self-dual gravitational field configuration \cite{2}, \cite{8}.

The following notations are used below: $\mk\subset\mK$ means a finite sub-complex  containing the centre of instanton with the boundary $\partial\mk\approx S^3$; ${\mK}'\subset\mK$ is an extra-large but finite sub-complex with the boundary $\partial{\mK}'\approx S^3$ containing the centre of instanton and vertexes $a_{\cV}\in{\mK}', \ {\cV}=1,2,\ldots{\mN}'\gg1$, so that the long-wavelength limit is valid and the continuous solution (\ref{inst20})-(\ref{inst50}) approximates correctly the exact lattice solution in a wide vicinity of $\partial{\mK}'$; the  hypersurface $\partial{\mK}'$ is given by the equation $r=R=\Const\longrightarrow\infty$. The Euler characteristics $\chi({\mk})=\chi({\mK}')=
\chi({\mK})=1$.

It has been proved in \cite{2} that there
exists the solution of the following system of equations and boundary conditions:
\begin{gather}
\delta\mA_g/\delta\omega^{\alpha}_{(\pm){\cW}mi}=0, \quad
\delta\mA_g/\delta e^a_{{\cW}mi}=0,
\nonumber \\
\omega^{\alpha}_{(-){\cW}mi}=0 \longleftrightarrow \Omega_{(-){\cW}ij}=1,
\label{inst60}
\end{gather}
\begin{gather}
\frac{i\sigma^{\alpha}}{2}\omega^{\alpha}_{(+)}\longrightarrow U^{-1}\d U \quad \mbox{as} \quad r\longrightarrow\infty,
\label{inst70}
\end{gather}
where
\begin{gather}
U=\exp\left(-\frac{i\sigma^3}{2}\varphi\right)\exp\left(\frac{i\sigma^2}{2}\theta\right)
\exp\left(-\frac{i\sigma^3}{2}\psi\right),
\label{inst80}
\end{gather}
and 
\begin{gather}
\Omega_{(+){\cW}ij}=-1,    \quad s^4_{\cW}\in\mk,
\nonumber \\
e^a_{{\cV}_1{\cV}_2}=\phi^a_{{\cV}_2}-\phi^a_{{\cV}_1}, \quad a_{{\cV}_1}a_{{\cV}_2}\in\mk, \quad  a_{{\cV}_1}a_{{\cV}_2}\notin\partial\mk.
\label{inst90}
\end{gather}
on $\mk$ ($a_{{\cV}_1}a_{{\cV}_2}$ is 1-simplex). The solution of Eqs. (\ref{inst60})-(\ref{inst90}) is lattice analogue of  Eguchi-Hanson self-dual solution. It is denoted as $\Omega_{(\mbox{inst}){\cV}_1{\cV}_2}, \ \   e^a_{(\mbox{inst}){\cV}_1{\cV}_2}$.

\section{The asymptotic behavior of fermion zero mode associated with gravitational instanton}

To begin with, we define the lattice variant of the (right) neutrino action. For that purpose
it is necessary to extract from the quantity (\ref{discr60}) the part interacting with the field
$\Omega_{(+){\cW}ij}$ only \footnote{The same procedure for constructing the neutrino action  imbedded in
gravity with Minkowski signature is valid.}:
\begin{gather}
\mA_{(+)}=-\frac{1}{5\cdot6\cdot24}\sum_{\cW}\sum_{i,j,k,l,m}\varepsilon_{{\cW}ijklm}\varepsilon_{abcd}\times
\nonumber \\
\times\Theta^a_{(+){\cW}mi}
e^b_{{\cW}mj}e^c_{{\cW}mk}e^d_{{\cW}ml},
\nonumber \\
\Theta^a_{(+){\cW}ij}=
\frac{1}{2}\left(\eta^{\dag}_{{\cW}i}\sigma^a
\Omega_{(+){\cW}ij}\phi_{{\cW}j}+\right.
\nonumber \\
\left.+\phi^{\dag}_{{\cW}j}\Omega_{(+){\cW}ji}(\sigma^a)^{\dag}\eta_{{\cW}i}\right).
\label{ferm1}
\end{gather}
It is convenient to write the continuous variant of the introduced fermion lattice action (\ref{ferm1})
in the form
\begin{gather}
\mA_{(+)}=\int\d^{(4)}x\,|\left(\det e^b_{\lambda}\right)|{}\left\{\frac{1}{2}e^{\mu}_a\left[
\eta^{\dag}\sigma^a\,{\cal D}_{(+)\mu}\,\phi+c.c.
\right]\right\}{},
\nonumber \\
{\cal D}_{(+)\mu}\equiv \partial_{\mu}+\frac{i}{2}\sigma^{\alpha}\omega^{\alpha}_{(+)\mu}.
\label{ferm10}
\end{gather}
The set of independent fermion variables is described by $\{\phi,\,\eta,\,\phi^{\dag},\,\eta^{\dag}\}$.

The actions (\ref{ferm1}), (\ref{ferm10}) can be interpreted as the lattice and continuous variants of (right)
neutrino actions, correspondingly.

Further  it is believed that the gravitational  fields in (\ref{ferm1})  are the lattice instanton solutions
(\ref{inst60})-(\ref{inst80}), and the gravitational  fields in (\ref{ferm10})  are the
corresponding fields in the long-wavelength limit (\ref{inst20})-(\ref{inst50}).

At the limit $r\longrightarrow\infty$ we have $g=1$. Let's introduce the designations for the case $g=1$:
\begin{gather}
\frac{i}{2}\sigma^{\alpha}\omega^{\alpha(0)}_{(+)\mu}\equiv\frac{i}{2}\sigma^{\alpha}\omega^{\alpha}_{(+)\mu}\big|_{g=1}=
\frac{i}{2}\bigg(
\varsigma^1\sigma^1+\varsigma^2\sigma^2+\varsigma^3\sigma^3 \bigg)_{\mu},
\nonumber \\
{\cal D}^{(0)}_{(+)\mu}\equiv \partial_{\mu}+\frac{i}{2}\sigma^{\alpha}\omega^{\alpha(0)}_{(+)\mu}.
\nonumber
\end{gather}
So we have
\begin{gather}
{\cal D}_{(+)\mu}={\cal D}^{(0)}_{(+)\mu}+
\frac{i}{2}
\bigg(
(1-g)\varsigma^1\sigma^1+(1-g)\varsigma^2\sigma^2+
\nonumber \\
+(1-g^2)\varsigma^3\sigma^3\bigg)_{\mu}.
\label{ferm30}
\end{gather}
It is easy to see that (the definition of $U\in\SU(2)$ is given in (\ref{inst80}))
\begin{gather}
{\cal D}^{(0)}_{(+)\mu}=U^{-1}\hat{\partial}_{\mu} U.
\label{ferm40}
\end{gather}
Combining Eqs. (\ref{ferm30}) and (\ref{ferm40}) we rewrite the Dirac-Weyl operator in
(\ref{ferm10}) as follows:
\begin{widetext}
\begin{gather}
\sigma^ae^{\mu}_a{\cal D}_{(+)\mu}=U^{-1}\left(U\sigma^aU^{-1}\right)e^{\mu}_a
\left\{\partial_{\mu}+\frac{i}{2}U\bigg(
(1-g)\varsigma^1\sigma^1+(1-g)\varsigma^2\sigma^2+(1-g^2)\varsigma^3\sigma^3\bigg)_{\mu}U^{-1}
\right\}U.
\label{ferm50}
\end{gather}
One can obtain the following row matrix
\begin{gather}
U\sigma^aU^{-1}\equiv\sigma^bA_b^a=
\nonumber \\
=(\sigma^1,\,\sigma^2,\,\sigma^3,\,i)
\left(
\begin{array}{cccc}
(\cos\theta\cos\varphi\cos\psi-\sin\varphi\sin\psi)  &   -(\cos\theta\cos\varphi\sin\psi+\sin\varphi\cos\psi)
& -\sin\theta\cos\varphi & 0 \\
(\cos\theta\sin\varphi\cos\psi+\cos\varphi\sin\psi)  &  (-\cos\theta\sin\varphi\sin\psi+\cos\varphi\cos\psi)
& -\sin\theta\sin\varphi & 0  \\
\sin\theta\cos\psi & -\sin\theta\sin\psi  &   \cos\theta &  0 \\
0  &  0  & 0  &  1  \\
\end{array} \right).
\label{ferm60}
\end{gather}
as a result of direct calculations. Further, according to Eqs. (\ref{inst30}) and (\ref{ferm60})
\begin{gather}
A_b^ae^{\mu}_a=
2\left(
\begin{array}{cccc}
-r^{-1}\sin\varphi \ \ &   -r^{-1}\cot\theta\cos\varphi  \ \
& (r\sin\theta)^{-1}\cos\varphi\left(\cos^2\theta+g^{-1}\sin^2\theta\right) \ \ & 0 \\
r^{-1}\cos\varphi  &   -r^{-1}\cot\theta\sin\varphi
& (r\sin\theta)^{-1}\sin\varphi\left(\cos^2\theta+g^{-1}\sin^2\theta\right) & 0  \\
0 & -r^{-1}  &   r^{-1}\left(1-g^{-1}\right)\cos\theta &  0 \\
0  &  0  & 0  &  g/2  \\
\end{array} \right).
\label{ferm70}
\end{gather}
Using the aforesaid formulae we transform the operator (\ref{ferm50}) into the form
\begin{gather}
\sigma^ae^{\mu}_a{\cal D}_{(+)\mu}=U^{-1}i\left\{\frac{2}{r}
 \left(
\begin{array}{cc}
-l_3 & l_-  \\
l_+ & l_3 \\
\end{array} \right)+g\frac{\partial}{\partial r}\right.
\nonumber \\
\left.+\left[\frac{2}{r\sin\theta}\left(\cos^2\theta+g^{-1}\sin^2\theta\right)
 \left(
\begin{array}{cc}
0 & e^{-i\varphi}  \\
e^{i\varphi} & 0 \\
\end{array} \right)
+\frac{2i}{r}\left(1-g^{-1}\right)\cos\theta
\left(
\begin{array}{cc}
1 & 0  \\
0 & -1 \\
\end{array}
\right) \right]\left(-i\frac{\partial}{\partial\psi}\right)+\frac{1+2g-3g^2}{rg}
 \right\}U,
\label{ferm80}
\end{gather}
\begin{gather}
 l_3=-i\frac{\partial}{\partial\varphi},  \qquad l_{\pm}=e^{\pm i\varphi}\left(\pm\frac{\partial}{\partial\theta}
 +i\cot\theta\frac{\partial}{\partial\varphi}\right),
\nonumber \\
[l_{\pm},\,l_3]=\mp l_{\pm}, \qquad [l_+,\,l_-]=2l_3.
\label{ferm90}
\end{gather}
\end{widetext}
We see that the operator in curly brackets in (\ref{ferm80}) does not depend on the variable $\psi$.
Therefore, it is  naturally to take the simplest ansatz for zero mode in the form
\begin{gather}
\phi=U^{-1}\tilde{\phi}, \quad \eta=U^{-1}\tilde{\eta},
\nonumber \\
(\partial/\partial\psi)\tilde{\phi}=0,
\quad (\partial/\partial\psi)\tilde{\eta}=0.
\label{ferm100}
\end{gather}
Thus  the operator (\ref{ferm80}) for the zero mode problem for $r\gg a$ reduces effectively
to the following one:
\begin{gather}
\sigma^ae^{\mu}_a{\cal D}_{(+)\mu}=U^{-1}i\left[\frac{2}{r}
 \left(
\begin{array}{cc}
-l_3 & l_-  \\
l_+ & l_3 \\
\end{array} \right)+\frac{\partial}{\partial r}\right]U,
\label{ferm110}
\end{gather}
and
the effective action describing fermion zero mode configuration takes the form
\begin{gather}
\mA_{(+)}=\int r^3\sin\theta\d r\d \theta\d\varphi\d\psi\times
\nonumber \\
\times\left\{\frac{i}{2}\tilde{\eta}^{\dag}
\left[\frac{2}{r}
 \left(
\begin{array}{cc}
-l_3 & l_-  \\
l_+ & l_3 \\
\end{array} \right)+\frac{\partial}{\partial r}\right]\tilde{\phi}+c.c.\right\}
\label{ferm115}
\end{gather}
because of
\begin{gather}
|\left(\det e^b_{\lambda}\right)|\d^{(4)}x=r^3\sin\theta\d r\d \theta\d\varphi\d\psi
\nonumber \\
\end{gather}
for instanton field solution.

Note that the frequently used operator $2{\bf ls}$ in hydrogen atom physics has the form
\begin{gather}
2{\bf ls}=\left(
\begin{array}{cc}
l_3 & l_-  \\
l_+ & -l_3 \\
\end{array} \right),
\nonumber
\end{gather}
it differs from that in Eq. (\ref{ferm115}).

The action stationarity condition relative to variable $\tilde{\eta}^{\dag}$  gives the zero mode equation
\begin{gather}
\left\{\frac{2}{r}
 \left(
\begin{array}{cc}
-l_3 & l_-  \\
l_+ & l_3 \\
\end{array} \right)+\frac{\partial}{\partial r}\right\}\tilde{\phi}_0=0.
\label{ferm120}
\end{gather}
The stationarity condition of the action (\ref{ferm115}) relative to variable $\tilde{\phi}$ yields
\begin{gather}
\frac{2}{r}
 \left(
\begin{array}{cc}
-l_3 & l_-  \\
l_+ & l_3 \\
\end{array} \right)\tilde{\eta}_0=\left(\frac{3}{r}+\frac{\partial}{\partial r}\right)\tilde{\eta}_0.
\label{ferm130}
\end{gather}

Eqs. (\ref{ferm120}) and (\ref{ferm130}) imply that  the function $\tilde{\phi}_0$ and $\tilde{\eta}_0$ are the eigenfunctions of the operator
\begin{gather}
\left(
\begin{array}{cc}
-l_3 & l_-  \\
l_+ & l_3 \\
\end{array} \right)
\label{ferm150}
\end{gather}
with the common eigenvalue $\lambda$.  Otherwise, the action (\ref{ferm115}) would be equal to zero identically
since the operator (\ref{ferm150}) is  Hermitean.

Let's consider the anzats
\begin{gather}
\tilde{\phi}_0=f(r)\left[\exp\left(-\frac{i\sigma^3}{2}\varphi\right)
\left(
\begin{array}{c}
h(\theta)/\sqrt{\sin\theta}  \\
k(\theta)/\sqrt{\sin\theta}  \\
\end{array} \right)\right].
\label{ferm160}
\end{gather}
Equation
\begin{gather}
\left(
\begin{array}{cc}
-l_3 & l_-  \\
l_+ & l_3 \\
\end{array} \right)\left[\exp\left(-\frac{i\sigma^3}{2}\varphi\right)
\left(
\begin{array}{c}
h(\theta)/\sqrt{\sin\theta}  \\
k(\theta)/\sqrt{\sin\theta}  \\
\end{array} \right)\right]=
\nonumber \\
=\lambda\left[\exp\left(-\frac{i\sigma^3}{2}\varphi\right)
\left(
\begin{array}{c}
h(\theta)/\sqrt{\sin\theta}  \\
k(\theta)/\sqrt{\sin\theta}  \\
\end{array} \right)\right]
\label{ferm170}
\end{gather}
is satisfied when and only when
\begin{gather}
\frac{\d k}{\d\theta}=-\left(\lambda-\frac12\right)h, \quad  \frac{\d h}{\d\theta}=\left(\lambda-\frac12\right)k.
\label{ferm21000}
\end{gather}

To have the acceptable boundary conditions at $\theta=0,\pi$, one must consider only
the eigenvalues $\lambda=(n+1/2), \ n=0,\pm1,\ldots$. Eq. (\ref{ferm120}) shows that the eigenvalues are acceptable
only for $n\geq1$. Otherwise the mode $\phi_0$ would be non-normalizable. On the other hand, the function $\eta_0$ would be $\O\left(r^{2(n-1)}\right)$ as $r\longrightarrow\infty$ for $n\geq2$ according to Eq. (\ref{ferm130}), i.e. it would be non-normalizable. Therefore the only acceptable eigenvalue is $\lambda=3/2$. Then Eqs. (\ref{ferm120}) and (\ref{ferm130}) give
\begin{gather}
\left(\frac{\d}{\d r}+\frac{3}{r}\right)f=0 \quad\longrightarrow \quad f\sim\frac{\Const}{r^3},
\label{ferm22000}
\end{gather}
\begin{gather}
\frac{\partial}{\partial r}\tilde{\eta}_0=0 \quad \mbox{as} \quad r\longrightarrow \infty.
\label{ferm2000}
\end{gather}

There are only two solutions:
\begin{gather}
\left(
\begin{array}{c}
h  \\
k  \\
\end{array} \right)^{(1)}=\sqrt{2}
\left(
\begin{array}{c}
\sin\theta  \\
\cos\theta  \\
\end{array} \right),    \quad
\left(
\begin{array}{c}
h  \\
k  \\
\end{array} \right)^{(2)}=\sqrt{2}
\left(
\begin{array}{c}
\cos\theta  \\
-\sin\theta  \\
\end{array} \right).
\label{ferm2100}
\end{gather}
\begin{widetext}
for $\lambda=3/2$. Combining Eqs. (\ref{inst80}), (\ref{ferm100}), (\ref{ferm160}), (\ref{ferm22000}), (\ref{ferm2000})  and (\ref{ferm2100}), we obtain two asymptotic solutions:
\begin{gather}
\phi_0^{(1)}=\frac{\Const}{r^3}\exp\left(\frac{i\sigma^3}{2}\psi\right)
\left(
\begin{array}{c}
\sqrt{\tan\left(\theta/2\right)}  \\
\sqrt{\cot\left(\theta/2\right)}  \\
\end{array} \right),
\quad
\eta_0^{(1)}=\Const\cdot\exp\left(\frac{i\sigma^3}{2}\psi\right)
\left(
\begin{array}{c}
\sqrt{\tan\left(\theta/2\right)}  \\
\sqrt{\cot\left(\theta/2\right)}  \\
\end{array} \right),
\nonumber \\
\phi_0^{(2)}=\frac{\Const}{r^3}\exp\left(\frac{i\sigma^3}{2}\psi\right)
\left(
\begin{array}{c}
\sqrt{\cot\left(\theta/2\right)}  \\
-\sqrt{\tan\left(\theta/2\right)}  \\
\end{array} \right) \quad,
\eta_0^{(2)}=\Const\cdot\exp\left(\frac{i\sigma^3}{2}\psi\right)
\left(
\begin{array}{c}
\sqrt{\cot\left(\theta/2\right)}  \\
-\sqrt{\tan\left(\theta/2\right)}  \\
\end{array} \right).
\label{ferm2200}
\end{gather}
\end{widetext}

It is known that two spinors $\phi$ and $\left(i\sigma^2\phi\right)^*$ transform identically
under the gauge transformations $\Spin(4)_{(+)}$ \cite{9}. Here the upper index ${}^*$ means complex conjugation.
But we have
\begin{gather}
\phi_0^{(2)}=\left(i\sigma^2\phi_0^{(1)}\right)^*.
\nonumber
\end{gather}
This equality leads to the conclusion that there is only one independent smooth fermion zero mode associated
with lattice gravitational instanton. Therefore any linear combination of the solutions (\ref{ferm2200})
can be considered as asymptotic behaviour of zero mode.

It will be proved that the corresponding lattice solution is normalizable.

Note that there is a great number of other solutions of Eq.
$\sigma^ae^{\mu}_a{\cal D}_{\mbox{inst}(+)\mu}\phi_0=0$.
We give an example of a series of the operator (\ref{ferm170}) eigenfunctions with eigenvalues $\lambda\neq3/2$:
\begin{gather}
\phi_0'=f'(r)\exp\left(\frac{i\sigma^3}{2}\psi\right)\left[e^{im\varphi}(\sin\theta)^m
\left(
\begin{array}{c}
\sqrt{\cot\left(\theta/2\right)}  \\
-\sqrt{\tan\left(\theta/2\right)}  \\
\end{array} \right)\right],
\nonumber \\
\lambda=\left(m+3/2\right), \quad m=1,\,2,\ldots
\nonumber
\end{gather}
As was shown above, the eigenvalues $\lambda\neq3/2$ are not acceptable.

\section{Existence of lattice fermion zero modes}

We must solve lattice equations
\begin{gather}
\delta\mA_{(+)}/\delta\phi_{\cV}=0, \quad \delta\mA_{(+)}/\delta\eta^{\dag}_{\cV}=0
\label{ferzer10}
\end{gather}
as well as their complex conjugate equations \footnote{The complex conjugate equations does not need to be considered
apparently.} for the action (\ref{ferm1}) taken on self-dual gravitational solution (\ref{inst60})-(\ref{inst90})
$\Omega_{(\mbox{inst}){\cV}_1{\cV}_2}, \ \   e^a_{(\mbox{inst}){\cV}_1{\cV}_2}$
with the boundary conditions (\ref{ferm2200}) as $r\longrightarrow\infty$.

In order to solve the problem, we use the method which has been successful in solving lattice pure gravity self-dual equations with given boundary conditions \cite{2}. The method can be applied to lattice theory, but it is
fundamentally unacceptable in the case of continuous theories. The reason is that the number of variables
(degrees of freedom) associated with finite space-time volume is finite in any lattice theory, while
{\it{the number of variables per volume is infinite (uncountable) in continuous theories}}.

Introduce the following Lagrange function on ${\mK}'$ (see the text between Eqs.
(\ref{inst50}) and (\ref{inst60})) depending on the variables $\{\phi_{\cV},\,\eta_{\cV},\,\phi^{\dag}_{\cV},\,\eta^{\dag}_{\cV}\},\, {\cV}=1,\ldots,\,{\mN}'$:
\begin{gather}
{\cal{L}}=-\frac{1}{5\cdot6\cdot24}\sum_{\cW:\,s^4_{\cW}\in{\mK}'}\sum_{i,j,k,l,m}\varepsilon_{{\cW}ijklm}
\varepsilon_{abcd}\times
\nonumber \\
\times\Theta^a_{(+){\cW}mi}
e^b_{{\cW}mj}e^c_{{\cW}mk}e^d_{{\cW}ml}-
\nonumber \\
-\lambda^{(\phi)}\Phi^{(\phi)}-\lambda^{(\eta)}\Phi^{(\eta)}-
\nonumber \\
-\left\{\sum_{{\cV}:\,a_{\cV}\in\partial{\mK}'}\sum_{s=1,2}\left(\lambda^{(\phi)}_{{\cV},s}\Phi^{(\phi)}_{{\cV},s}+\lambda^{(\eta)}_{{\cV},s}\Phi^{(\eta)}_{{\cV},s}\right)
+c.c.\right\}.
\label{ferzer20}
\end{gather}
Here $\{\lambda\}$ are  Lagrange multipliers.

The constraints
\begin{gather}
\Phi^{(\phi)}_{{\cV},s}=\left(\frac{\phi^s_{\cV}}{\phi_0^{s}(x_{\cV})}-
\frac{\phi^s_{{\cV}_0}}{\phi_0^{s}(x_{{\cV}_0)})}\right),
\nonumber \\
\Phi^{(\eta)}_{{\cV},s}=\left(\frac{\eta^s_{\cV}}{\eta_0^{s}(x_{\cV})}-
\frac{\eta^s_{{\cV}_0}}{\eta_0^{s}(x_{{\cV}_0)})}\right),
\nonumber \\
a_{\cV}\in\partial{\mK}', \quad a_{{\cV}_0}\notin\partial{\mK}'.
\label{ferzer30}
\end{gather}
fix the boundary conditions (\ref{ferm2200}) near the hypersurface $\partial{\mK}'$. Here $x_{\cV}$ and $x_{{\cV}_0}$ are the coordinate values at the vertexes $a_{\cV}$ and
$a_{{\cV}_0}$ correspondingly (see (\ref{discr80}));  $a_{{\cV}_0}$ is a fixed vertex from immediate neighborhood
of hypersurface $\partial{\mK}'$, so that there is a 1-simplex $a_{{\cV}_0}a_{\cV}$ for some vertex
$a_{\cV}\in\partial{\mK}'$; index $s=1,2$
enumerates the components of Weyl spinors $\phi^{\dag},\,\phi$.

The constraints
\begin{gather}
\Phi^{(\phi)}=\left(\sum_{{\cV}:\,a_{\cV}\in({\mK}'\setminus\partial{\mK}')}v_{\cV}\sum_{s=1,2}|\phi^s_{\cV}|^2-1\right),
\nonumber \\
\Phi^{(\eta)}=\left(\sum_{{\cV}:\,a_{\cV}\in({\mK}'\setminus\partial{\mK}')}v_{\cV}\sum_{s=1,2}|\eta^s_{\cV}|^2-1\right)
\label{ferzer40}
\end{gather}
mean that the fermion field configurations are normalizable on ${\mK}'$,
\begin{gather}
v_{\cV}=\frac15\sum_{{\cW}:\,a_{\cV}\in{\cW}}v_{\cW},
\label{ferzer50}
\end{gather}
\begin{gather}
v_{\cW}=\frac{1}{4!}\varepsilon_{{\cW}ijklm}
e^1_{{\cW}mi}e^2_{{\cW}mj}e^3_{{\cW}mk}e^4_{{\cW}ml}.
\label{ferzer60}
\end{gather}
The expression (\ref{ferzer60}) means the oriented volume of the 4-simplex $s_{\cW}$,
factor $1/4!$ is required since the volume of a
four-dimensional parallelepiped with generatrices
$e^1_{{\cW}mi},\;e^2_{{\cW}mj},\;e^3_{{\cW}mk}$, and $e^4_{{\cW}ml}$ is
$4!$ times larger than the volume of a 4-simplex with the same
generatrices. The expression  $v_{\cV}$ in (\ref{ferzer50}) is the
sum of the volumes $v_{\cW}$ for that ${\cW}$-4-simplexes which
contain the vertex $a_{\cV}$, the factor $1/5$ is necessary due to the fact that all five
vertices of each simplex are taken into account independently in (\ref{ferzer40}). So,
the volume (\ref{ferzer50}) is the specific volume per vertex. In the
long-wavelength limit the constraints (\ref{ferzer40}) transform into
(the same is true for $\Phi^{(\eta)}$)
\begin{gather}
\Phi^{(\phi)}=\left(\int_{{\mK}'}\left(\sum_{s=1,2}|\phi^s(x)|^2\right)\, e^1\wedge e^2\wedge e^3\wedge e^4-1\right),
\nonumber \\
e^a=e^a_{\mu}\d x^{\mu}.
\nonumber
\end{gather}

Since the sub-complex ${\mK}'$ contains a finite number ${\mN}'$ of vertexes, so the Lagrange function
(\ref{ferzer20}) depends on a finite number of classical variables $\{\phi^{\dag}_{\cV},\,\phi_{\cV}\},\, {\cV}=1,\ldots,\,{\mN}'<\infty$. For "not patologic" complexes $\mK$ the estimation
\begin{gather}
{\mN}'\sim R^4
\label{ferzer70}
\end{gather}
is valid.

The problem is as follows: the local maximums and minimums of Lagrange function (\ref{ferzer20})
 constrained by the constraints (\ref{ferzer30}) and (\ref{ferzer40}) are to be founded.
The simplicity of the constraints simplifies very much the problem: the constraints can be solved
evidently. Thus the constraints (\ref{ferzer30}) give
\begin{gather}
\phi^s_{\cV}=\left(\frac{\phi_0^{s}(x_{\cV})}{\phi_0^{s}(x_{{\cV}_0)})}\right)\phi^s_{{\cV}_0},  \quad
\eta^s_{\cV}=\left(\frac{\eta_0^{s}(x_{\cV})}{\eta_0^{s}(x_{{\cV}_0)})}\right)\eta^s_{{\cV}_0},
\nonumber \\
a_{\cV}\in\partial{\mK}',
\quad a_{{\cV}_0}\notin\partial{\mK}'.
\label{ferzer80}
\end{gather}
It is useful to divide the Lagrange function (\ref{ferzer20}) into two terms:
\begin{gather}
{\cal{L}}={\cal{L}}'+\partial{\cal{L}}.
\label{ferzer85}
\end{gather}
Here ${\cal{L}}'$ does not depend on the variables $(\phi_{{\cV}_0},\,\eta_{{\cV}_0},\,
\phi^{\dag}_{{\cV}_0},\,\eta^{\dag}_{{\cV}_0})$, while $\partial{\cal{L}}$ is a homogeneous linear form
for these variables. Evidently, $\partial{\cal{L}}$ depends only on the variables associated
with vertexes  from immediate neighborhood of hypersurface $\partial{\mK}'$.

Realization of the constraints (\ref{ferzer40}) converts the Lagrange function (\ref{ferzer20})  into
a smooth function defined on compact metric finite-dimensional manifold ${\cal{C}}$ without boundary.
It is well known for this case that the Lagrange function is a bounded one and it has the local maximum(s) and minimum(s) at some points $p_{\xi}\in{\cal{C}}$.  Moreover, since the space ${\cal{C}}$ is without boundary, so the total differentials of the Lagrange function at the points $p_{\xi}$ are equal to zero.

It should be emphasized that the total differential of the Lagrange function must be calculated
with respect to independent variables. The variables associated with the vertexes
$a_{\cV}\in\partial{\mK}'$ are expressed evidently in terms of independent variables $\phi_{{\cV}_0},\,
\eta_{{\cV}_0}$ according to  Eqs. (\ref{ferzer80}). Let $a_{{\cV}_1}\notin\partial{\mK}'$ be
a fixed vertex from immediate neighborhood of hypersurface $\partial{\mK}'$. The constraints
(\ref{ferzer40}) will be resolved if we express, for example, the real component of $\phi^1_{{\cV}_1}$ and
$\eta^1_{{\cV}_1}$ in terms of the rest independent variables:
\begin{gather}
\Re\,\phi^1_{{\cV}_1}=\pm\frac{1}{\sqrt{v_{{\cV}_1}}}\sqrt{1-
\sum_{{\cV}:\,a_{\cV}\in({\mK}'\setminus\partial{\mK}'),\,s=1,2}^{\ \ \ \ \ \prime}v_{\cV}|\phi^s_{\cV}|^2},
\label{ferzer90}
\end{gather}
and analogously for $\Re\,\eta^1_{{\cV}_1}$. Here the prime above the sum means that the variable $\Re\,\phi^1_{{\cV}_1}$
is absent. Thus
\begin{gather}
\frac{\partial\Re\,\phi^1_{{\cV}_1}}{\partial\phi^s_{\cV}}=\mp\left(\frac{v_{\cV}}{v_{{\cV}_1}}\right)
\frac{\phi^s_{\cV}}{\Re\,\phi^1_{{\cV}_1}}.
\label{ferzer100}
\end{gather}
Therefore one should replace
\begin{gather}
\frac{\partial}{\partial\phi_{{\cV}}} \longrightarrow \frac{\partial}{\partial\phi_{{\cV}}}\mp\left[\left(\frac{v_{\cV}}{v_{{\cV}_1}}\right)
\frac{(\phi_{\cV})^*}{\Re\,\phi^1_{{\cV}_1}}\right]\frac{\partial}{\partial\Re\,\phi^1_{{\cV}_1}},
\label{ferzer110}
\end{gather}
and so on.

Let's consider the stationarity condition for the Lagrange function (Lagrange multipliers can be put equal to zero)
relative to the variable $\eta^{\dag}_{\cV}$:
\begin{gather}
\frac{\partial{\cal L}}{\partial\eta^{\dag}_{{\cV}}}\mp\left[\left(\frac{v_{\cV}}{v_{{\cV}_1}}\right)
\frac{\eta_{\cV}}{\Re\,(\eta^{\dag})^1_{{\cV}_1}}\right]\frac{\partial{\cal L}}{\partial\Re\,(\eta^{\dag})^1_{{\cV}_1}}=0.
\label{ferzer120}
\end{gather}
For ${\cV}={\cV}_0$ we have the same equation, but it is convenient to divide the Lagrange function
according to (\ref{ferzer85}):
\begin{gather}
\frac{\partial{\cal L}'}{\partial\eta^{\dag}_{{\cV}_0}}+\frac{\partial(\partial{\cal L})}{\partial\eta^{\dag}_{{\cV}_0}}
\mp\left[\left(\frac{v_{\cV}}{v_{{\cV}_1}}\right)
\frac{\eta_{{\cV}_0}}{\Re\,(\eta^{\dag})^1_{{\cV}_1}}\right]\frac{\partial{\cal L}}{\partial\Re\,(\eta^{\dag})^1_{{\cV}_1}}=0.
\label{ferzer130}
\end{gather}

Now pass to the limit $R\longrightarrow\infty$ in Eqs. (\ref{ferzer120})-(\ref{ferzer130}).

We have
\begin{gather}
 v_{\cV}/v_{{\cV}_1}\sim1.
\label{ferzer140}
\end{gather}
for "not patologic" complex. There is also the estimation
\begin{gather}
\frac{\eta_{\cV}}{\Re\,(\eta^{\dag})^1_{{\cV}_1}}\sim\O\left(R^0\right)
\label{ferzer150}
\end{gather}
as a consequence of the boundary condition (\ref{ferm2000}). Finally, the estimation
\begin{gather}
\frac{\partial{\cal L}}{\partial\Re\,(\eta^{\dag})^1_{{\cV}_1}}\sim\O\left(R^{-3}\right)
\label{ferzer160}
\end{gather}
is true since the quantity $\partial{\cal L}/\partial\Re\,(\eta^{\dag})^1_{{\cV}_1}$ depends linearly only on
a limited number (of the order of one) of the variables $\phi_{\cV}$ and due to the boundary condition
(\ref{ferm22000}).

The estimation of the second term in the left hand side of Eq. (\ref{ferzer130}) can be obtained
if we take into account Eqs. (\ref{ferzer80}) and definition of the quantity $\partial{\cal L}$
(see Eq. (\ref{ferzer85})). The angular dependence of the boundary
variables is defined according to (\ref{ferzer80}). Therefore, this quantity resides in stationary point relative to the angular variations by definition of $\partial{\cal L}$.
So the derivative $\partial/\partial\eta^{\dag}_{{\cV}_0}$
comes to the derivative with respect to  $r$  acting into variables $\phi_{\cV}$ for  vertexes
$a_{\cV}$ from the immediate neighborhood of hypersurface $\partial{\mK}'$, and
 $\partial{\cal L}$ is the sum of the quantities which are of the order of $\O\left(R^{-4}\right)$,
but the number of these quantities (the number of the vertexes on $\partial{\mK}'$) is of the order of $R^3$.
Thus
\begin{gather}
\frac{\partial(\partial{\cal L})}{\partial\eta^{\dag}_{{\cV}_0}}\sim\O\left(R^{-1}\right).
\label{ferzer170}
\end{gather}
Using the estimations (\ref{ferzer160}) and (\ref{ferzer170}) we obtain estimations
\begin{gather}
\frac{\partial{\cal L}}{\partial\eta^{\dag}_{{\cV}}}\sim\O\left(R^{-3}\right),  \quad
\frac{\partial{\cal L}}{\partial\eta^{\dag}_{{\cV}_0}}\sim\O\left(R^{-1}\right).
\label{ferzer180}
\end{gather}

From (\ref{ferm2000}) and (\ref{ferzer40}) it follows that
\begin{gather}
\eta^{\dag}_{{\cV}}\sim\O\left(R^{-2}\right).
\label{ferzer190}
\end{gather}
As a result of estimations (\ref{ferzer180}) and (\ref{ferzer190}) we obtain:
\begin{gather}
{\cal L}'\sim\sum_{{\cV}:\,a_{\cV}\in({\mK}'\setminus\partial{\mK}'),\,{\cV}\neq{\cV}_0}
\eta^{\dag}_{{\cV}}\frac{\partial{\cal L}}{\partial\eta^{\dag}_{{\cV}}}\sim
\nonumber \\
\sim\O\left(R^{4}\right)\cdot\O\left(R^{-2}\right)\cdot\O\left(R^{-3}\right)\sim\O\left(R^{-1}\right).
\label{ferzer200}
\end{gather}

Now one should consider the stationary point of the Lagrange function (\ref{ferzer20}) relative to the variables
$\phi_{\cV}$.  For this purpose it's enough to make a replacement $\eta^{\dag}_{\cV}\longrightarrow\phi_{\cV}$
in Eqs. (\ref{ferzer120})-(\ref{ferzer130}).  The estimation
\begin{gather}
\frac{\partial{\cal L}}{\partial\phi_{\cV}}\sim\O\left(R^{-2}\right)
\label{ferzer210}
\end{gather}
is true since the quantity $\partial{\cal L}/\partial\Re\,\phi^1_{{\cV}_1}$ depends linearly only on a limited number (of the order of one) of the variables $\eta^{\dag}_{\cV}$ and due to the
estimation (\ref{ferzer190}). Further, since the derivative $\left(\partial\eta^{\dag}_{\cV}/\partial r\right)$
is negligibly small  in the neighborhood of hypersurface $\partial{\mK}'$ (see (\ref{ferm2000})),
so the quantity
\begin{gather}
\frac{\partial(\partial{\cal L})}{\partial\phi_{{\cV}_0}}
\nonumber
\end{gather}
is also negligibly small. Therefore the estimation (\ref{ferzer210}) is valid for all
$a_{\cV}\in{\mK}'\setminus\partial{\mK}'$.

Now one should pass to the limit $R\longrightarrow\infty$. It follows from  the
 estimations
(\ref{ferzer180}) and (\ref{ferzer210}) that the problem (\ref{ferzer10})-(\ref{ferzer20})  possess a solution. Besides,
according to (\ref{ferzer200}) the action (\ref{ferm1}) is equal to zero on this solution.
This means that the discussed solution really is a  fermion zero mode.

\section{Discussion: the number of lattice zero modes and  their nature}

The important question is unanswered:  how many lattice solutions for zero modes do exist?
We can not prove rigorously here that there exist two linearly independent lattice zero modes, but
we formulate the conjecture:  \\

{\it {Hypothesis}}. There exist two linearly independent lattice fermion zero modes associated
with lattice gravitational instanton.  One of them possesses the properties of a usual smooth mode,
another can be characterized as a singular mode \cite{1}. \\

We give here only some reasoning to justify the hypothesis.

Suppose that the  sub-complex $\mk$ (centre of instanton) is large enough, i.e. the number
of 4-simplexes $s^4_{\cW}\in\mk$ is a large number. According to (\ref{ferm1}) and (\ref{inst90})
the contribution to the fermion action which is associated with sub-complex $\mk$ is
\begin{gather}
\mA_{(+)(\mk)}=\frac{1}{2\cdot5\cdot6\cdot24}\sum_{{\cW}:\,s^4_{\cW}\in\mk}\sum_{i,j,k,l,m}\varepsilon_{{\cW}ijklm}\varepsilon_{abcd}\times
\nonumber \\
\times\left(\eta^{\dag}_{{\cW}m}\sigma^a
\phi_{{\cW}i}+c.c.\right)e^b_{{\cW}mj}e^c_{{\cW}mk}e^d_{{\cW}ml}.
\label{ferzer220}
\end{gather}
Let's consider two adjacent 4-simplexes $s^4_{\cW}=a_{{\cW}i}a_{{\cW}j}a_{{\cW}k}a_{{\cW}l}a_{{\cW}m}$
and $s^4_{{\cW}'}=a_{{\cW}'i'}a_{{\cW}j}a_{{\cW}k}a_{{\cW}l}a_{{\cW}m}$ with common 3-simplex
$s^3=a_{{\cW}j}a_{{\cW}k}a_{{\cW}l}a_{{\cW}m}$ and  different vertexes $a_{{\cW}i}$ and $a_{{\cW}'i'}$.
Evidently, $s^4_{\cW}$ and $s^4_{{\cW}'}$ have the opposite orientations. Therefore
\begin{gather}
\varepsilon_{{\cW}ijklm}=-\varepsilon_{{\cW}'i'jklm}.
\label{ferzer230}
\end{gather}
This implies that  the contribution into (\ref{ferzer220})
associated with 3-simplex $s^3=a_{{\cW}j}a_{{\cW}k}a_{{\cW}l}a_{{\cW}m}$ vanishes for $\phi_{{\cW}i}=\phi_{{\cW}'i'}$. Note
that there is no cavities in $\mK$ (and hence in $\mk$) by definition. This means that each
3-simplex $s^3=a_{{\cW}j}a_{{\cW}k}a_{{\cW}l}a_{{\cW}m}\in\left(\mk\setminus\partial\mk\right)$ belongs to two and only two
adjacent 4-simplexes $s^4_{\cW}=a_{{\cW}i}a_{{\cW}j}a_{{\cW}k}a_{{\cW}l}a_{{\cW}m}\in\mk$
and $s^4_{{\cW}'}=a_{{\cW}'i'}a_{{\cW}j}a_{{\cW}k}a_{{\cW}l}a_{{\cW}m}\in\mk$.   It follows from this consideration that
the contribution (\ref{ferzer220}) vanishes on the configuration
\begin{gather}
\phi_{\cV}=\Const \quad \mbox{on} \quad \mk\setminus\partial\mk.
\label{ferzer240}
\end{gather}
In other words, the configuration (\ref{ferzer240})
satisfies Eq. (\ref{ferzer10}) on $\mk\setminus\partial\mk$.

This consideration leads to the hypothesis that the configuration (\ref{ferzer240}) is a
part of the configuration of a regular zero mode on the instanton interior.

According to the Hypothesis the irregular zero mode (doubled fermion in the Wilson sense)
does exist also.

The Hypothesis is proved mathematically rigorously in the case of the Dirac zero modes for the Yang-Mills smooth instantons.
The idea of the proof is based on the fact that the normal smooth fermion modes give the known
anomaly contribution into the chiral current. But a trivial consequence of our definition of lattice Dirac
fermions is the fact that the lattice fermion measure does not contain an axial anomaly. This means that the Dirac irregular modes
compensate completely the contribution of the smooth fermion modes into anomaly. Since in the Yang-Mills
theory the Dirac zero modes and anomal contributions into axial current are inextricably connected
(concerning the smooth modes the statement is demonstrated in Introduction),  both normal and anomal zero modes must exist. The detailed calculations are given in \cite{1}.

The problem formulated here as a Hypothesis requires a detailed study, as well as the physical consequences
of the fermion zero mode existence.

As a final matter we give some comments regarding the Wilson fermion doubling problem.
It is well known that the lattice Dirac fermions possessing the chiral symmetry property possess also
the  Wilson fermion doubling property. The statement is valid for regular lattices \cite{10,11} as well as for
irregular lattices (simplicial complexes) \cite{1}.  However there is a qualitative difference between the
phenomena on regular and irregular lattices. In the case of regular lattice there are 16 doublers,
and all of the quanta of all doublers propagate identically like free particles (in free theory). But there is a qualitative difference between the dynamics of soft regular and soft irregular quanta in the case of irregular lattice.  While the regular quanta propagate as free particles since they have lost the information about lattice, the irregular doubled quanta can not propagate in the space-time for the reason that
the irregular quanta wave functions are determined essentially by irregular "breathing" lattice \footnote{We use the term
"breathing" lattice in the considered theory since the variables describing the lattice are the dynamic variables
and they fluctuate strongly in quantum theory.}. Therefore, the irregular quanta can not be observed directly,
but only by means of some physical effects taking place due to the existence of irregular quanta (more detailed
comments on the question is contained in \cite{1}). In a sense the irregular quanta are not observable
since they are not relevant for the most part of physics. We note also that the number 16 for the doublers for the
cubic lattice is related with the cubic symmetry. Since general irregular lattice (simplicial complex) have no symmetries,
the irregular quanta enumeration problem remains unsolved.

Note also that the zero modes differ from soft modes qualitatively: zero mode is
localized in the vicinity of instanton and annihilates the Dirac operator precisely, while soft mode
is an eigen-mode of the Dirac operator with non-zero eigenvalue.

\begin{acknowledgments}

 The work has been supported by the RScF grant 16-12-10151.

\end{acknowledgments}

\end{document}